\magnification=1200 \vsize=8.9truein \hsize=6.5truein \baselineskip=0.6truecm
\parindent=0truecm \parskip=0.2truecm
\nopagenumbers \font\scap=cmcsc10 \hfuzz=0.8truecm
\font\tenmsb=msbm10
\font\sevenmsb=msbm7
\font\fivemsb=msbm5
\newfam\msbfam
\textfont\msbfam=\tenmsb
\scriptfont\msbfam=\sevenmsb 
\scriptscriptfont\msbfam=\fivemsb
\def\Bbb#1{{\fam\msbfam\relax#1}}

\def\xdo{x_{n-1}}
\def\xup{x_{n+1}}

\def\zdo{z_{n-1}}
\def\zup{z_{n+1}}
\def\ado{a_{n-1}}
\def\aup{a_{n+1}}
\def\bdo{b_{n-1}}
\def\bup{b_{n+1}}

\def\sc{singularity confinement }

\null \bigskip  \centerline{\bf Singularity  confinement and algebraic entropy: }
\smallskip  \centerline{\bf the case of the discrete Painlev\'e equations }

\vskip 2truecm
\bigskip
\centerline{\scap Y. Ohta}
\centerline{\sl Department of Applied Mathematics}
\centerline{\sl Faculty of Engineering, Hiroshima University}
\centerline{\sl 1-4-1 Kagamiyama, Higashi-Hiroshima 739-8527, Japan}

\bigskip
\centerline{\scap K. M. Tamizhmani}
\centerline{\sl Department of Mathematics}
\centerline{\sl Pondicherry University}
\centerline{\sl Kalapet, Pondicherry, 605014 India}
\bigskip
\centerline{\scap B. Grammaticos}
\centerline{\sl GMPIB, Universit\'e Paris VII}
\centerline{\sl Tour 24-14, 5$^e$ \'etage}
\centerline{\sl 75251 Paris, France}
\bigskip
\centerline{\scap A. Ramani}
\centerline{\sl CPT, Ecole Polytechnique}
\centerline{\sl CNRS, UMR 7644}
\centerline{\sl 91128 Palaiseau, France}

Abstract
\medskip
We examine the validity of the results obtained with the \sc integrability criterion in the case of discrete Painlev\'e
equations. The method used is based on the requirement of non-exponential growth of the homogeneous degree of the iterate of
the mapping. We show that when we start from an integrable autonomous mapping and deautonomise it using \sc the degrees of
growth of the nonautonomous mapping  and of the autonomous one are {\sl identical}. Thus this
low-growth based approach is compatible with the integrability of the results obtained through singularity confinement. 
The origin of the \sc property and its necessary character for integrability are also analysed. 

\vfill\eject
\footline={\hfill\folio} \pageno=2

The \sc property has been proposed some years ago [1] as a discrete  integrability criterion. The essence
of the method is the observation that in integrable mappings a spontaneously appearing singularity
does not propagate {\sl ad infinitum} under the action of the mapping but disappears (``is confined'')
after some iteration steps. Thus \sc appeared as a necessary condition for discrete integrability. However
the sufficiency of the criterion was not unambiguously established. The attitude (of the present authors
at least) has always been  that if the \sc condition were strong enough  then it would suffice for
integrablity, in perfect analogy with the Painlev\'e-ARS [2] property for
continuous systems. This sufficiency of the \sc criterion was recently challenged by Hietarinta and Viallet [3] who produced
explicit examples of  mappings satisfying \sc  which are {\sl not} integrable to the point of
exhibiting chaotic behaviour. Their approach is based on the relation of discrete integrability and
the complexity of the evolution introduced by Arnold and Veselov. According to Arnold [4] the
complexity (in the case of mappings of the plane) is the number of intersection points of a fixed curve with the image of a
second curve obtained under the mapping at hand. While the complexity grows exponentially with the iteration for generic
mappings, it can be shown [5] to grow only polynomially for a large class of integrable mappings. As Veselov
points out,  ``integrability has an essential correlation with the weak growth of certain characteristics''.
Thus the authors of [3]  proposed to directly test the degree of the successive iterates
and introduced the notion of algebraic entropy. 
The method is appropriate for birational mappings. One starts by introducing homogeneous coordinates and
studies the degree of the iterate. As Bellon and Viallet [6] remark, the growth of the degree is invariant under coordinate
changes though the degree itself is not.  A generic (non integrable) mapping leads to degrees that grow exponentially. The
algebraic entropy is thus naturally defined as $E=\lim_{n\to \infty} \log(d_n)/n$ where $d_n$ is the degree of the $n$-th
iterate. Thus nonintegrable mappings have nonzero  algebraic entropy. The conjecture in [3,6] is that integrability implies
polynomial growth, leading to zero algebraic entropy.
 
The main application of the \sc approach was the derivation and study of discrete Painlev\'e equations (d-$\Bbb P$'s) [7]. On the
light of the results of Hietarinta and Viallet which have shown that the criterion used was not restrictive enough, one
might be tempted to doubt the integrability of the mappings obtained (despite a considerable volume of integrability-confirming
results). The aim of this paper is to show that these doubts are unjustified and to confirm the validity of the
approach previously used, with the help of algebraic entropy techniques.

Let us first recall what has always been our approach to the derivation of d-$\Bbb P$'s. We
start from an autonomous system the integrability of which has been independently established. In the case of d-$\Bbb P$'s,
this system is the QRT mapping [8]:
$$ f^{(1)}(x_n)-(\xup+\xdo)f^{(2)}(x_n)+\xup\xdo f^{(3)}(x_n)=0\eqno(1)$$
When the $f^{(i)}$'s are quartic functions, satisfying specific
constraints, the mapping (1) is integrable in terms of elliptic functions. Since the elliptic functions are the
autonomous limits of the Painlev\'e transcendents, the mapping (1) is the appropriate starting point for the construction of the
nonautonomous discrete systems which are the analogues of the Painlev\'e equations. The procedure we used, often referred to
as `deautonomisation', consists in finding the dependence of the coefficients of the quartic polynomials
appearing in (1) with respect to the independent variable $n$, which is compatible with the \sc property. Namely, the
$n$-dependence is obtained by asking that the singularities are indeed confined. One rule that has always been used, albeit
often tacitly, is that confinement must be implemented ``the soonest possible''. What this rule really means is that the
singularity pattern of the deautonomised mapping must be the same as the one of the autonomous mapping. Our claim is that a
deautonomisation with a different singularity pattern (for instance a `later' confinement) would lead to a non-integrable
system.  The reason why this deautonomisation procedure can be justified is the following. Since the autonomous starting point
is integrable, it is expected that the growth of the degree of the iterates is polynomial. Now it turns out that the
application of the \sc deautonomisation corresponds to the requirement that the nonautonomous mappings lead to the same
factorizations and subsequent simplifications and have precisely the same growth properties as the autonomous ones.  These
considerations will be made more transparent thanks to the examples we present in what follows.

Let us start with a simple case. We consider the mapping:
$$\xup+\xdo={ax_n+b\over x_n^2} \eqno(2)$$
where $a$ and $b$ are constants. In order to compute the degree of the iterates we introduce homogeneous coordinates by taking
$x_0$=$p$,   $x_1$=$q/r$, assuming that the degree of $p$ is zero, and compute the degree of homogeneity in $q$ and $r$  at
every iteration. We could have of course introduced a different choice for $x_0$ but it turns out that
the choice of a zero-degree $x_0$ considerably simplifies the calculations.  We obtain thus the degrees: 0, 1, 2, 5,
8, 13, 18, 25, 32, 41, \dots, . Clearly the degree growth is polynomial. We have
$d_{2m}=2m^2$ and $d_{2m+1}=2m^2+2m+1$. This is in perfect agreement with the fact that the
mapping (2) is integrable (in terms of elliptic functions), being a member of the QRT family of integrable mappings.
(A remark is necessary at this point. In order to
obtain a closed-form expression for the degrees of the iterates, we start by computing a sufficient number of them. Once the
expression of the degree has been heuristically established we compute the next few ones and check that they agree
with the analytical expression predicted). 
We now turn to the deautonomisation of the mapping. The \sc result is that $a$ and $b$ must satisfy the conditions
 $\aup-2a_n+\ado=0$, $\bup=\bdo$, i.e. $a$ is linear in $n$ while $b$ is a constant with an even/odd dependence.
Assuming now that $a$ and $b$ are arbitrary functions of $n$ we compute the degrees of the iterates of (2). We obtain
successively 0, 1, 2, 5, 10, 21, 42, 85,\dots. The growth is now exponential, the degrees behaving like 
$d_{2m-1}=(2^{2m}-1)/3$  and $d_{2m}=2d_{2m-1}$, a clear indication that the mapping is not integrable in
general.  Already at the fourth iteration the  degrees differ in the autonomous and nonautonomous cases. 
Our approach consists in requiring that the degree in the nonautonomous case be {\sl identical} to the one obtained in the
autonomous one. If we implement the requirement that $d_4$ be 8 instead of 10 we find two conditions $\aup-2a_n+\ado=0$,
$\bup=\bdo$, i.e. precisely the ones obtained through singularity confinement. Moreover, once these two conditions are
satisfied, the subsequent degrees of the nonautonomous case coincide with that of the autonomous one. Thus this mapping, leading
to polynomial growth, should be integrable, and, in fact, it is. As we have shown in [9], where we presented its Lax pair,
equation (2) with $a(n)=\alpha n+\beta$ and $b$ constant (the even-odd dependence can be gauged out by a parity-dependent
rescaling of the variable $x$) is a discrete form of the Painlev\'e I equation. In the examples that follow, we shall show
that in all cases the nonautonomous form of an integrable mapping obtained through \sc leads to exactly the same degrees of
the iterates as the autonomous one. 

Our second example is a multiplicative mapping:
$$\xup\xdo={a_nx_n+b\over x_n^2} \eqno(3)$$ 
where one can put $b=1$ through an appropriate gauge. In the autonomous case we obtain, starting with $x_0$=$p$ and
$x_1$=$q/r$, successively the degrees: 0, 1, 2, 3, 4, 7, 10, 13, 16, 21, 26, \dots, i.e. again a quadratic growth. In fact, if
$n$ is of the form $4m+k$, ($k$=0,1,2,3) the degree is given by $d_n=4m^2+(2m+1)k$. The deautonomisation of (3) is
straightforward. We compute the successive degrees and find: 0, 1, 2, 3, 4, 7, 11, \dots, . At this stage we require that a
factorization occur in order to bring the degree $d_6$ from 11 to 10. The condition for this is $a_{n+2}a_{n-2}=a_n^2$ i.e. $a$
of the form
$a_{e,o}\lambda_{e,o}^n$ with an even-odd dependence which can be easily gauged away. This condition is sufficient in order to
bring the degrees of the successive iterates down to the values obtained in the autonomous case. Quite expectedly the
condition on $a$ is {\sl precisely} the one obtained by singularity confinement. The Lax pair of (3) can be easily obtained
from our results in [10]. We find  that if we introduce the matrices:
$L_n=\left(\matrix {0&0&k\over x_n&0\cr 0&0&\xdo&q\xdo\cr hx_n&0&1&q\cr 0&{hk_{n-1}\over \xdo}&0&0} \right)$ and 
$M_n=\left(\matrix{0&x_n\over k(x_n+1)&0&0\cr 0&0&1&0\cr 0&0&{1\over x_n}&{q\over x_n}\cr h&0&0&0} \right)$ we can obtain from
the compatibility $L_{n+1}M_n(h/q)=M_n(h)L_n$ the equation 
$\xup\xdo=k_nk_{n+1}(x_n+1)/x_n^2$, where $k_{n+1}=qk_{n-1}$, which is equivalent to (3) up to a gauge transformation.

The case of the mapping
$$\xup+\xdo=a_n+{b_n\over x_n} \eqno(4)$$
has a more interesting deautonomisation. In this case we make a slightly different choice of homogeneous coordinates, which
simplifies the results for the degrees of the iterates. We assume $x_0=p/r$, $x_1$=$q/r$ and compute the degree of homogeneity
in $p$, $q$ and $r$. We find $d_n$=1, 1, 2, 3, 5, 8, 11, 15, 20, 25, 31, 38, \dots, i.e. if $n$ is of the form 3$m$ we have
$d_n$=$3m^2-m+1$, for $n$=3$m$+1, $d_n$=$3m^2+m+1$, and for $n$=3$m$+2, $d_n$=$3m^2+3m+2$. In the generic nonautonomous case
the corresponding degrees are 1, 1, 2, 3, 5, 8, 13, \dots,. The requirement that $d_6$=11 leads to the condition $\aup=\ado$
and $b_{n+2}-\bup-\bdo+b_{n-2}$=0. Thus $b$ is linear with a {\sl ternary} symmetry while $a$ is a
constant (with an even/odd dependence which can be gauged away). This fully nonautonomous form of (4) is a discrete
form of Painlev\'e IV studied in [11] and [12] where we have given its Lax pair.

We now turn to what is known as the ``standard'' discrete Painlev\'e equations [7] and compare the results of \sc to those of the
algebraic entropy approach. We start with d-P$_{\rm I}$ in the form:
$$\xup+x_n+\xdo=a_n+{b_n\over x_n} \eqno(5)$$
The degrees of the iterates of the autonomous mapping are 0, 1, 2, 3, 6, 9, 12, 17, 22, \dots, i.e. a quadratic growth with
$d_{3m+k}$=$3m^2+(2m+1)k$, for $k=$0,1,2 while those of the
generic  nonautonomous one are 0, 1, 2, 3, 6, 11, \dots, . Requiring two extra factorisations at that level (so as to bring $d_5$ down
to 9) we find the following conditions $\aup=a_n$, so $a$ must be a constant, and $b_{n+2}-\bup-b_n+\bdo$=0, i.e. $b$ is of
the form $b_n=\alpha n+\beta+\gamma (-1)^n$ which are exactly the result of singularity confinement.
Implementing these conditions we find that the autonomous and nonautonomous mappings have the same (polynomial) growth [6]. Both
are integrable, the Lax pair of the nonautonomous one, namely d-P$_{\rm I}$ having been given in [10,13,14].

For the discrete P$_{\rm II}$ equation we have
$$\xup+\xdo={a_nx_n+b_n\over x_n^2-1} \eqno(6)$$
The degrees of the iterates in the autonomous case are $d_n$=0, 1, 2, 4, 6, 9, 12, 16, 20, \dots, (i.e. $d_{2m-1}$=$m^2$,
$d_{2m}$=$m^2+m$) while in the generic nonautonomous case we find the first discrepancy for $d_4$ which is now 8. To
bring it down to 6 we find two conditions, $\aup-2a_n+\ado$=0 and $\bup=\bdo$. This means that $a$ is linear in $n$ and $b$ is
an even/odd constant, as predicted by singularity confinement. Once we implement these constraints, the degrees of the
nonautonomous and autonomous cases coincide. The Lax pair of equation (6) in the nonautonomous form, i.e. 
d-P$_{\rm II}$, has been presented in [10,15,16].  

The $q$-P$_{\rm III}$ equation was obtained from the deautonomisation of the mapping:
$$\xup\xdo={(x_n-a_n)(x_n-b_n)\over (1-c_nx_n)(1-x_n/c_n)} \eqno(7)$$
In the autonomous case we obtain the degrees $d_n$=0, 1, 2, 5, 8, 13, 18, \dots, just like for equation (2), while in  the
generic nonautonomous case we have 0, 1, 2, 5, 12,\dots, .  For $d_4$ to be 8 instead of 12, one needs four factors to cancel
out. The conditions are $c_{n+1}=c_{n-1}$ and $\aup\bdo=\ado\bup=a_nb_n$. Thus $c$ is a constant up to an  even/odd
dependence, while $a$ and $b$ are proportional to $\lambda^n$ for some $\lambda$, with an extra even/odd dependence, just as
predicted by
\sc in [7]. The Lax pair for $q$-P$_{\rm III}$ has been presented in [10,17].  

For the remaining three discrete Painlev\'e  equations the Lax pairs are not known yet. It is thus important to have one more
check of their integrability provided by the algebraic entropy approach. We start with d-P$_{\rm IV}$ in the form:
$$(\xup+x_n)(\xdo+x_n)={(x_n^2-a^2)(x_n^2-b^2)\over (x_n+z_n)^2-c^2} \eqno(8)$$
where $a$, $b$ and $c$ are constants. If $z_n$ is constant we obtain for the degrees of the successive iterates $d_n$=0, 1, 3, 6,
11, 17, 24, \dots, . The general expression of the growth is $d_n$=6$m^2$ if $n=3m$, $d_n$=6$m^2+4m+1$ if $n=3m+1$ and
$d_n$=6$m^2+8m+3$ if $n=3m+2$.  This polynomial (quadratic) growth is expected since in the autonomous case this equation is
integrable, its solution being given in terms of elliptic functions. For a generic
$z_n$ we obtain the sequence  $d_n$=0, 1, 3, 6, 13,
\dots, . The condition for the extra factorizations to occur in the last case, bringing down the degree $d_4$ to 11, is for $z$
to be linear in $n$. We can check that the subsequent degrees coincide with those of the autonomous case. 
 
For the $q$-P$_{\rm V}$ we start from:
$$(\xup x_n-1)(\xdo x_n-1)={(x_n^2+ax_n+1)(x_n^2+bx_n+1)\over (1-z_ncx_n)(1-z_ndx_n)} \eqno(9)$$
where $a$, $b$, $c$ and $d$ are constants. If moreover $z$ is also a constant, we obtain exactly the same sequence  of degrees 
$d_n$=0, 1, 3, 6, 11, 17, 24, \dots, as in the d-P$_{\rm IV}$ case.  Again, this polynomial (quadratic) growth is expected
since this mapping is also integrable in terms of elliptic functions. For the generic nonautonomous case we again find the
sequence $d_n$=0, 1, 3, 6, 13, \dots, . Once more we require a factorization  bringing down $d_4$ to 11. It turns out that this
entails a $z$ which is exponential in $n$, which then generates the same sequence of degrees as the autonomous case. In both the
d-P$_{\rm IV}$ and $q$-P$_{\rm V}$ cases we find the $n$-dependence already obtained through singularity confinement. Since
this results to a vanishing algebraic entropy we expect both equations to be integrable.

The final system we shall study is the one related to the discrete P$_{\rm VI}$ equation:
$${(\xup x_n-\zup z_n)(\xdo x_n-\zdo z_n)\over (\xup x_n-1)(\xdo x_n-1)}={(x_n^2+az_nx_n+z_n^2)(x_n^2+bz_nx_n+z_n^2)\over
(x_n^2+cx_n+1)(x_n^2+dx_n+1)}\eqno(10)$$ where $a$, $b$, $c$ and $d$ are constants. In fact the generic symmetric
QRT mapping can be brought to the  autonomous ($z_n$ constant) form of equation (10) through the appropriate homographic
transformation.  In the autonomous case, we obtain the degree sequence
$d_n$=0, 1, 4, 9, 16, 25, \dots, i.e. $d_n$=$n^2$. Since mapping (10) is rather complicated we cannot investigate its full
freedom. Still we were able to perform two interesting calculations. First, assume that  in the rhs instead of the function $z_n$
a different function $\zeta_n$ appears. In this case the degrees grow like 0, 1, 5, \dots, and the condition to have $d_2$=4
instead of 5 is
$\zup\zdo z_n^2=\zeta_n^4$. Assuming this is true, we compute the degree $d_3$ of the next iterate and find $d_3$=13 instead of 9.
To bring down $d_3$ to the value 9 we need $z_n^2=\zeta_n^2$, which up to a redefinition of $a$ and $b$ means $z_n=\zeta_n$. This
implies $\zup\zdo=z_n^2$, and $z_n$ is thus an exponential function of $n$, $z_n$=$\lambda^n$ (which is in agreement with the
results of [18]). Then  a quartic factor drops out and $d_3$ is just 9. One can then check that the next degree is 16,
just as in the autonomous case. Thus the $q$-P$_{\rm VI}$ equation leads to the same growth as the generic symmetric QRT mapping
and is thus expected to be integrable.
As a matter of fact we were able to show that the generic {\sl asymmetric} QRT mapping leads to the same growth $d_n$=$n^2$ as
the symmetric one. This is not surprising, given the integrability of this mapping. 
What is interesting is that the growth of the generic  symmetric and asymmetric QRT mappings are the same. Thus $d_n$=$n^2$ is
the maximal growth one can obtain for the QRT mapping in the homogeneous variables we are using. As a matter of fact we have
also checked that the asymmetric nonautonomous $q$-P$_{\rm VI}$ equation, introduced in [18] led to exactly the same degree
growth $d_n$=$n^2$.

Let us summarize our findings. In this paper, we have compared the method of \sc and the approach based on the study of
algebraic entropy when applied to the deautonomisation of integrable mappings. We have shown that in every case the confinement
condition which ensured that the singularity pattern of the autonomous and non-autonomous cases are identical was precisely the
one necessary in order to bring the growth down to the one obtained in the autonomous case. This validates the deautonomisation
results obtained through
\sc at least in the domain of d-$\Bbb P$'s. This suggests also a strategy for the study of integrable mappings. We believe that
in the light of the present results, when one starts from an integrable autonomous mapping, the deautonomisation can be
performed solely with the help of singularity confinement, a procedure considerably simpler than the calculation of the
algebraic entropy. 

Our present investigation also sheds light on the singularity confinement, and its necessary character as discrete integrability
criterion. Let us go back to the example of mapping (2) with $b=1$. We start with $x_0=p$, $x_1=q/r$. Iterating further we find 
$$x_2={r^2+aqr-pq^2\over  q^2},\quad x_3={qP_4\over  r(r^2+aqr-pq^2)^2},\quad x_4={(r^2+aqr-pq^2)P_6\over  P_4^2}, \quad
x_5={P_4P_9\over rP_6^2}$$ 
where the $P_k$'s are homogeneous polynomials in $q$, $r$ of degree $k$. (Remember that $p$ is of zero homogeneous degree in our
convention).  The pattern now becomes clear. Whenever a new polynomial appears in the numerator of $x_n$ its square will appear
in the denominator of $x_{n+1}$ and it will appear one last time as a factor of the numerator of $x_{n+2}$, after which it
disappears due to factorisations. The singularities we are working with in the \sc approach correspond to the zeros of any of
these polynomials, which explains the pattern $\{0,\infty^2, 0\}$.
The \sc is intimately related to this factorisation which plays a crucial role in the algebraic entropy approach. 
Let us suppose now that $a$ is a generic function of $n$. In this case we get the sequence:
$$x_2={r^2+a_1qr-pq^2\over  q^2},\quad x_3={qQ_4\over  r(r^2+a_1qr-pq^2)^2},\quad x_4={(r^2+a_1qr-pq^2)Q_7\over qQ_4^2}$$
$$\quad x_5={qQ_4Q_{12}\over r(r^2+a_1qr-pq^2)Q_7^2}$$
where the $Q_k$'s are also homogeneous polynomials in $q$, $r$ of degree $k$. Now the simplifications that do occur are
insufficient to curb the asymptotic growth. As a matter of fact, if we follow a particular factor we can check that it keeps
appearing either in the numerator or the denominator (where its degree is alternatively 1 and 2). This corresponds to the
unconfined singularity pattern $\{0,\infty^2,0,\infty,0,\infty^2,0,\infty,\dots\}$. Once more, the confinement condition
$\aup-2a+\ado=0$ is the condition for $q$ to divide exactly $Q_7$, for both $q$ and $r^2+a_1qr-pq^2$ to divide exactly
$Q_{12}$, etc..
Our analysis clearly shows why \sc is necessary for integrability while not being sufficient in general. Still in the case of
integrable deautonomisation it does lead to the correct answer, which explains its success in the derivation of the discrete
Painlev\'e equations.

\noindent {\scap Acknowledgements}.
\smallskip
\noindent  
The financial help of the {\scap cefipra}, through the contract 1201-1, is gratefully acknowledged.
The authors are grateful to J. Fitch who provided them with a new (beta) version of {\scap reduce} without which the
calculations presented here would have been impossible. B. Grammaticos acknowledges interesting discussions with J.
Hietarinta.
\bigskip
{\scap References}
\smallskip
\item{[1]} B. Grammaticos, A. Ramani and V. Papageorgiou, Phys. Rev. Lett. 67 (1991) 1825.
\item{[2]} M.J. Ablowitz, A. Ramani and H. Segur, Lett. Nuov. Cim. 23 (1978) 333.
\item{[3]} J. Hietarinta and C.-M. Viallet, Phys. Rev. Lett. 81 (1998) 325.
\item{[4]} V.I. Arnold, Bol. Soc. Bras. Mat. 21 (1990) 1.
\item{[5]} A.P. Veselov, Comm. Math. Phys. 145 (1992) 181.
\item{[6]} M.P. Bellon and C.-M. Viallet, {\sl Algebraic Entropy}, Comm. Math. Phys. to appear.
\item{[7]} A. Ramani, B. Grammaticos and J. Hietarinta, Phys. Rev. Lett. 67 (1991) 1829.
\item{[8]} G.R.W. Quispel, J.A.G. Roberts and C.J. Thompson, Physica D34 (1989) 183. 
\item{[9]} A.S. Fokas, B. Grammaticos and A. Ramani, J. Math. An. and Appl. 180 (1993) 342.
\item{[10]} V.G. Papageorgiou, F.W. Nijhoff, B. Grammaticos and A. Ramani, Phys. Lett. A 164 (1992) 57.
\item{[11]} B. Grammaticos, A. Ramani, V. Papageorgiou, Phys. Lett. A 235 (1997) 475.
\item{[12]} B. Grammaticos, Y. Ohta, A. Ramani, H. Sakai, J. Phys. A 31 (1998) 3545.
\item{[13]} A.R. Its, A.V. Kitaev and A.S. Fokas, Usp. Mat. Nauk 45,6 (1990) 135.
\item{[14]} B. Grammaticos A. Ramani, J. Phys. A 31 (1998) 5787.
\item{[15]} N. Joshi, D. Bartonclay, R.G. Halburd, Lett. Math. Phys. 26 (1992) 123.
\item{[16]} B. Grammaticos, F.W. Nijhoff, V.G. Papageorgiou, A. Ramani and J. Satsuma, Phys. Lett. A185 (1994) 446.
\item{[17]} M. Jimbo and H. Sakai, Lett. Math. Phys. 38 (1996) 145. 
\item{[18]} B. Grammaticos and A. Ramani, Phys. Lett. A 257 (1999) 288.
\end